\documentclass[twocolumn, aps, pre, showpacs]{revtex4}
\usepackage{graphicx}
\usepackage{graphics}
\usepackage{amsmath}
\usepackage{bm}

\begin{document}

\title{Critical aging of a ferromagnetic system from a completely ordered state}
\author{Andrei A. Fedorenko}
\author{Steffen Trimper}

\affiliation{Fachbereich Physik, Martin-Luther-Universit\"{a}t Halle-Wittenberg,
D-06099 Halle, Germany}
\date{\today}

\begin{abstract}
We adapt the non-linear $\sigma$ model to study the nonequilibrium critical dynamics
of $O(n)$ symmetric  ferromagnetic system. Using the renormalization group analysis
in $d=2+\varepsilon$ dimensions we investigate the pure relaxation of the system
starting from a completely ordered state. We find that the average magnetization
obeys the long-time scaling behavior almost immediately after the system starts to evolve
while the correlation and response functions demonstrate scaling behavior which is
typical for aging phenomena. The corresponding fluctuation-dissipation ratio is
computed to first order in $\varepsilon$ and the relation between transverse and
longitudinal fluctuations is discussed.
\end{abstract}

\maketitle

Aging phenomena have been found in a broad variety of
strongly disordered systems such as polymer and spin glasses
\cite{hodge-95}, electronic system of an Anderson
insulator \cite{vaknin-00}, array of flux lines pinned by
disorder \cite{schehr-04}, etc. Recently much attention has been
attracted by aging of pure (or weakly disordered) systems with slow
relaxation dynamics governed, for example, by domain
growth \cite{bray-94} as in a ferromagnetic system below
the critical temperature $T_c$ or by critical slowing down as in
a ferromagnetic system exactly at criticality \cite{godreche-00}.
One expects that the critical aging phenomena can be cast into
different universality classes of nonequilibrium critical dynamics.
Most of studies consider the relaxation of a ferromagnetic system
starting from a completely disordered state after quenching it to the
fixed temperature $T\le T_c$. It was found that the response function
$R(t,s)$ and the correlation function $C(t,s)$ depend nontrivially on
the ratio $x=t/s$ similar to that found in glassy systems.
Here $s$ and $t$  are  waiting and observation times respectively.
The distance from equilibrium
can be measured by the fluctuation-dissipation ratio (FDR)
$X(t,s)= T R(t,s)/\partial_sC(t,s)$.
It has been argued that for critical aging the limit
$X^{\infty}=\lim_{t,s \to \infty} X(t,s)$ is a novel universal quantity of
critical phenomena \cite{godreche-00}. The FDR was computed for the
$d$ dimensional spherical model \cite{godreche-00},
$O(n)$ symmetric ferromagnetic model   \cite{calabrese-02} and diluted spin models
\cite{calabrese-02b,schehr-05,chen-05}.
In all these systems $X^{\infty}$ has values ranging between $0$ and $1/2$.
The mean field calculations show that the aging behavior is modified in the
presence of long-range correlations in the initial disordered state
\cite{newman-90,picone-02}.
Much less known about the relaxation starting from an ordered state.
The numerical simulations show that the correlation function
demonstrates behavior which is typical for aging phenomena
\cite{schehr-05,berthier-01},
while the magnetization obeys the long-time scaling behavior almost immediately after
the system starts to evolve \cite{jaster-99,li-96}.  This observation was used
to develop new effective numerical methods to determine the critical exponents,
which are based on the short-time critical dynamics and do not require the
time-consuming equilibration of the system \cite{zheng-96}.
However, up to now there is no
any theoretical explanation why the long-time scaling behavior
emerges already in the macroscopically early initial stage of relaxation
and there is no theoretical framework which allows one to take properly
into account the critical fluctuations in this regime of aging.

In this paper we study the nonequilibrium critical dynamics
of a ferromagnetic system starting from a completely ordered
state. The long-distance properties of the $O(n)$ symmetric system
below the transition point can be related to the non-linear $\sigma$
model defined by the reduced Hamiltonian \cite{polyakov-75}
\begin{equation}
\mathcal{H}=  \int d^d x \left[\frac1{2}(\nabla \mathbf{s})^2 -
\mathbf{ h}\cdot\mathbf{s}\right],
\hspace{1cm} |\mathbf{s}(x)|^2=1. \label{model}
\end{equation}
In principle, to describe properly the dynamics of a real isotropic magnet
one has to consider the Larmor precession of the conserved field $\mathbf{s}$
in the local magnetic field $\mathbf{h}$, \textit{i.e.} construct the
low-temperature version of the model~J~\cite{trimper-78}.
However in this work, which to our knowledge is the first analytical
study of the critical aging of a finite range system from a completely
ordered state, we are interested in the influence of the initial condition rather
than in the complications caused by more realistic dynamic model.
We restrict consideration to the pure relaxational dynamics (model A)
with the non-conserved $n$-component order parameter $\mathbf{s}(x)$.
It can be described by the Langevin equation
$
\partial_t \mathbf{s}= -\lambda {\delta \mathcal{H}}/{\delta \mathbf{s}}
+ \bm{\zeta},
$
where $\lambda$ is an Onsager coefficient and $\bm{\zeta}$ is a random thermal
noise \cite{dominicis-77,zinn-justin,bausch-80}.
Because of the constraint $\mathbf{s}^2=1$ the random noise can act only in
the tangential direction and thus is $\mathbf{s}$ - dependent. This dependence
complicates the explicit expression for thermal noise distribution,
however, one can simply formulate the dynamics in terms of a generating
functional \cite{bausch-80}. The corresponding weight reads
\begin{equation}
P[\mathbf{s},\hat{\mathbf{s}}]=\delta(\mathbf{s}^2-1)
\delta(\mathbf{s}\cdot\hat{\mathbf{s}}) \exp(-\mathcal{A}), \label{weight1}
\end{equation}
where $\hat{\mathbf{s}}$ is the response field and the action is given by
\begin{equation}
\mathcal{A}[\mathbf{s},\hat{\mathbf{s}}]=\frac1{T}\int d^dxdt \left(
\lambda \hat{\mathbf{s}}^2 + i \hat{\mathbf{s}}\cdot \left[
\partial_t \mathbf{s} + \lambda \frac{\delta \mathcal{H}}{\delta \mathbf{s}}
\right]\right).
\end{equation}
Note that $\hat{\mathbf{s}}$ is a real variable which is connected with the
corresponding quantity $\tilde{\mathbf{s}}$ introduced in
Refs.~\cite{calabrese-02,calabrese-02b,schehr-05,chen-05} by
$\tilde{\mathbf{s}} =i\hat{\mathbf{s}}$.
The first $\delta$-function in Eq.~(\ref{weight1}) is due to the constraint
$\mathbf{s}^2=1$ and the additional constraint $\mathbf{s}\cdot\hat{\mathbf{s}}=0$,
imposed by the second $\delta$-function, ensures that the thermal noise
$\bm{\zeta}$ acts only in the tangential direction \cite{bausch-80}.

The effect of a macroscopic initial condition  can be taken into
account by averaging over the  initial configurations
$\mathbf{s}_0(x)\equiv \mathbf{s}(t=0,x)$ with a weight
$P[\mathbf{s}_0]=\exp(-\mathcal{H}_0[\mathbf{s}_0]/T)$ \cite{janssen-89}.
Taking $\mathcal{H}_0=-\int d^dx\ \mathbf{h}_0 \cdot\mathbf{s}_0(x)$
we specify an initial condition that corresponds to the
equilibrium state of the system subjected to
an external magnetic field $\mathbf{h}_0$, which is switched off at $t=0$.
There is a deep analogy between the considered model and the renormalization
group (RG) description of the surface critical phenomena using expansion
about the low critical dimension \cite{diehl-86}.
We assume that the magnetic fields $\mathbf{h}_0$ and $\mathbf{h}$ act along the
one direction which we will refer as to longitudinal one.
We decompose the fields $\mathbf{s}=(\sigma,\bm{\pi})$ and
$\hat{\mathbf{s}}=(\hat{\sigma},\hat{\bm{\pi}})$
into $(n-1)$-component transverse parts $\bm{\pi}$ and $\hat{\bm{\pi}}$ and
longitudinal parts $\sigma=\sqrt{1-{\pi}^2}$ and
$\hat{\sigma}=-\bm{\pi}\cdot\hat{\bm{\pi}}/\sqrt{1-{\pi}^2}$,
where the constraints on the fields  have been used.
We now can write the weight functional
$\exp({-\cal{A}_{\mathrm{tr}}[\bm{\pi},\hat{\bm{\pi}}]})$
for the transverse components only with the action given by
\begin{eqnarray}
&&\!\! \mathcal{A}_{\mathrm{tr}}[\bm{\pi},\hat{\bm{\pi}}]=\frac1{T}\int d^dx \int_0^{\infty}
  dt \left\{\lambda \hat{\bm{\pi}}^2+i \hat{\bm{\pi}}(\partial_t-\lambda\nabla^2){\bm{\pi}}
 \right. \notag\\
&&\!\!
+\lambda\frac{(\bm{\pi}\cdot\hat{\bm{\pi}})^2}{1-\bm{\pi}^2}
+ \frac{i}{2} \frac{\bm{\pi}\cdot\hat{\bm{\pi}}}{1-\bm{\pi}^2}
\left[(\partial_t-\lambda\nabla^2){\bm{\pi}}^2-
\frac{\lambda}{2}\frac{ (\nabla{\bm{\pi}}^2)^2}{1-\bm{\pi}^2}\right. \notag \\
&& \left.\left.
+2\lambda h \sqrt{1-\bm{\pi}^2} \right]\right\}-
 \frac{h_0}{T} \int d^dx \sqrt {1-\bm{\pi}_0^2}. \label{Atr} \ \ \ \
\end{eqnarray}
We will adopt the dimensional regularization scheme \cite{zinn-justin} in which
the terms generated by the measure in the path integral over
$\mathbf{s}$ and $\hat{\mathbf{s}}$ vanish
so that we have omitted these terms in Eq.~(\ref{Atr}) from the beginning.
Let us introduce the connected Green functions
$
G_{k\hat{k}}^{l\hat{l}} := {\langle [\bm{\pi}]^{k}
[\hat{\bm{\pi}}]^{\hat{k}} [{\bm{\pi}}_0]^{l}
[\hat{\bm{\pi}}_0]^{\hat{l}} \rangle}
$,
the low temperature expansions of which can be obtained with a loop
expansion based on the action (\ref{Atr}).
The terms quadratic in $\bm{\pi}$ and $\hat{\bm{\pi}}$ give us the
free response function and the free correlator:
\begin{eqnarray}
 R^{0}_{q}(t,s) &=& -i  \Theta (t-s) G_q(t-s),
  \label{R1} \\
 C^{0}_{q}(t,s) &=&  \frac{G_q(|t-s|)}{h
+q^2} +  \left(h_{0}^{-1}-\frac{1}{h
+q^2} \right)  G_q(t+s),\ \ \ \ \label{C1}
\end{eqnarray}
where the notation $G_q(t):= T e^{-\lambda (h +q^2)t}$ is introduced.
The infinite number of higher order terms in the expansion of square
roots in Eq.~(\ref{Atr}) will be treated as interactions. In
each order in $T$ we have to take into account only a fixed number
of such terms. The completely ordered initial state corresponds to
the limit $h_0\to\infty$ in which Eq.~(\ref{C1}) becomes a Dirichlet correlator.
Although the action has infinite number of vertices proportional to $h_0$,
which are located at time surface $t=0$, it is easy to show by direct
inspection of Feynman diagrams that they do not contribute in the limit
$h_0\to \infty$. The finite $h_0^{-1}$  can be treated then as an additional
perturbation.
The great advantage of the considered nonequilibrium model in comparison with
the equilibrium counterpart is that the Green functions
are not spoiled  by  ir singularities,
so that we can probe the critical domain just taking the limit $h\to 0$
without using the less convenient matching procedure \cite{bausch-80,nelson-77}.
However, the theory suffers of the uv divergences which can be
converted into poles in $\varepsilon=d-2$ using dimensional regularization.
Exploiting the ideas of Refs.~\cite{polyakov-75} and \cite{zinn-justin} one can prove
the renormalizability of the model,
which means that all uv  divergences can be absorbed into finite number of
Z-factors according to
\begin{eqnarray}
&&\!\!\!\!\!\! \mathring{\bm{\pi}} = Z^{1/2}\bm{\pi},\ \
   \mathring{\hat{\bm{\pi}}} = \hat{Z}Z^{-1/2}\hat{\bm{\pi}}, \ \
  \mathring{\bm{\pi}}_0 = (ZZ_0)^{1/2}\bm{\pi}_0,\ \ \ \notag \\
&&\!\!\!\!\!\!  \mathring{\hat{\bm{\pi}}}_0 = \hat{Z}Z^{-1/2}Z_0^{1/2}\hat{\bm{\pi}}_0,
\ \
\mathring{\lambda}=Z\hat{Z}^{-1}\lambda, \ \ \mathring{T}=\mu^{-\varepsilon}K_dZ_TT,
  \notag \\
&& \!\!\!\!\!\! \mathring{h}=\mu^{2}Z_T Z^{-1/2}h, \ \
 \mathring{h}_0=\mu^{2}Z_T(ZZ_0)^{-1/2}h_0. \label{rencond}
\end{eqnarray}%
Here circles denote the bare quantities, $(2\pi)^dK_d$ is the surface
area of a $d$-dimensional  unit sphere, and $\mu$ is an
arbitrary momentum scale.
The Z-factors except for $Z_0$ are the same as in equilibrium
\cite{dominicis-77,zinn-justin} and  to order $T$ given by
$Z=\hat{Z}=1+(n-1)T/\varepsilon$ and $Z_T=1+(n-2)T/\varepsilon$. The
new factor $Z_0=1+(n-3)T/\varepsilon+O(T^2)$
serves to cancel the divergences arising
from the free correlator (\ref{C1}) for $t+s\to 0$.
The renormalized Green function reads
\begin{equation}
\mathring{G}_{k\hat{k}}^{l\hat{l}}=Z^{k/2}(\hat{Z}Z^{-1/2})^{\hat{k}}
(ZZ_0)^{l/2}(Z_0^{1/2} \hat{Z} Z^{-1/2})^{\hat{l}} G_{k \hat{k}}^{l\hat{l}}
\end{equation}
and satisfies the RG equation
\begin{eqnarray}
&&\!\!\!\!\!\!\!\!\!\!\!\!\! \left[ \mu {\partial}_{\mu} +\beta_T\partial_T
 +(\hat{\zeta}-\zeta)\lambda\partial_{\lambda}
+\rho h \partial_h + (\rho+\zeta_0/2) h_0\partial_{h_0} \right. \notag \\
&& \!\!\!\!\!\!\!\!\!\!\!\! +\left. (l+k-\hat{l}-\hat{k})\frac{\zeta}{2}
+(\hat{l}+\hat{k})\hat{\zeta}+
(l+\hat{l})\frac{\zeta_0}{2} \right]G_{k \hat{k}}^{l \hat{l}} = 0,
\label{RGeq}
\end{eqnarray}
with $\beta_{T}=\left. \mu {\partial}_{\mu} T \right|_{0}$,
$\zeta_{i}=\left. \mu {\partial}_{\mu} \ln Z_i \right|_{0}$,
$\rho=\beta_T/T+\zeta/2-d$,  where  $\left. \right|_{0}$ denotes
the derivative at fixed bare parameters. Note that the equation similar
to Eq.~(\ref{RGeq}) holds  for longitudinal and mixed Green functions.

The critical behavior of the non-linear $\sigma$ model is controlled by the
unstable ir fixed point (FP). For $n>2$ and $d>2$ the beta function has a
nontrivial zero $T_c=\varepsilon/(n-2)+O(\varepsilon^2)$,
which is related to the bare critical temperature $\mathring{T}_c$.
The solution of Eq.~(\ref{RGeq}), in conjunction with the simple
dimensional analysis, yields the scaling behavior of the Green function
at FP as
\begin{eqnarray}
&&\!\!\!\!\!\!\!\! G_{k \hat{k}}^{l \hat{l}}(q,t;T,h,h_0) = \xi(T)^{-d_G}M(T)^{l+k-\hat{l}-\hat{k}}
\hat{M}(T)^{\hat{l}+\hat{k}}  \notag \\
&&\ \times M_0(T)^{l+\hat{l}} F_{k \hat{k}}^{l \hat{l}}
(q\xi,t\xi^{-z};hM\xi^{d}/T,h_0 M_0/h)
\label{scaling}
\end{eqnarray}
where $\xi(T)=\mu^{-1}T^{1/\varepsilon}%
 \exp(\int_0^T dT^{\prime}(1/\beta_T(T^{\prime})-1/\varepsilon T^{\prime}))$
is the correlation length and
$d_G=d(k+\hat{k}+l+\hat{l}-1)-2(\hat{l}+\hat{k})$  the
canonical dimension of $G_{k \hat{k}}^{l \hat{l}}$.
The scaling functions $M$, $\hat{M}$ and $M_0$
are given by
$\ln M_i(T)=-\frac12 \int_0^T dT^{\prime}\zeta_i(T^{\prime})/\beta_T(T^{\prime})$.
Note that in statics $M$ has a meaning of the spontaneous magnetization.
For $\tau=T-T_c \to 0^{-}$ we derive $\xi\sim|\tau|^{-\nu}$ and  $M\sim|\tau|^{\beta}$,
with critical exponents $\nu=-1/\beta_T(T_c)$ and $\beta=\nu\zeta(T_c)/2$. The dynamic
exponent reads $z=2+\hat{\zeta}(T_c)-\zeta(T_c)$. To one loop order we have
$\nu=1/\varepsilon$, $\beta=(n-1)/2(n-2)$ and $z=2+O(\varepsilon^2)$.

We now focus on the scaling properties of two-times quantities
which describe the critical system evolving from a completely
ordered state. To that end we put in what follows $h=h_0^{-1}=0$
and  apply  to Eq.~(\ref{scaling}) a short-time expansion of
the fields $\bm{\pi}$ and $\hat{\bm{\pi}}$ in terms of the initial
fields $\bm{\pi}_0$ and $\hat{\bm{\pi}}_0$
\cite{janssen-89,fedorenko-04}. As a result we obtain the scaling
behavior of the response and correlation functions for
$s\xi^{-z}\to 0$ and $t \xi^{-z}>0$:
\begin{eqnarray}
&& \!\!\!\!\!\!\! R_q(t,s) = A_R s^{-(z-2+\eta)/z}( t/s)^{\bar{\theta}}
   f_R (\lambda q^z(t-s),t/s ), \ \ \ \ \label{scale1} \\
&& \!\!\!\!\!\!\! C_q(t,s) = A_C s^{(2-\eta)/z}( t/s)^{\bar{\theta}}f_C
   (\lambda q^z(t-s),t/s), \label{scale2}
\end{eqnarray}
where $\eta=2\beta/\nu-\varepsilon$ is the Fisher exponent.
The new dynamic critical exponent is defined as
$ \bar{\theta}=(2-\eta-z-\zeta_0(T_c)/2)/z$, and should be distinguished
from the initial slip exponent $\theta$ \cite{janssen-89}.
To lowest order in $\varepsilon$ it is given by
$\bar{\theta}=-\varepsilon(n-1)/4(n-2)+O(\varepsilon^2)$
and coincides with $-\beta/\nu z$. Although at this point we do not
see any symmetry which ensures this identity, we argue that it
holds to all orders in $\varepsilon$ and below give some arguments
supporting this conjecture.
The functions $f_R$ and  $f_C$ are regular functions of both
arguments and finite for $s \to 0$.
We have explicitly computed the renormalized
response and correlation functions to one-loop order:
\begin{eqnarray}
 R_q(t,s) &=& K_d^{-1}\Omega[t,s] R^0_q(t,s)
 \label{rel0}, \ \  \\
 C_q(t,s) &=& K_d^{-1}\Omega[t,s] C^0_q(t,s) \notag \\
&+&  {(n-2)T^2}e^{-\lambda q^2 (t+s)}
 F(2\lambda q^2 s)/{2K_d q^2}, \ \ \  \  \
 \label{rel1}
\end{eqnarray}
where we have introduced notations
$\Omega[t,s]=1-\frac{T}{2}(\gamma+\ln2\lambda\mu^2s%
 +\frac{n-1}{2}\ln\frac{t}{s} )$  and
 $F(y)=\gamma+\ln|y|- \mathrm{Ei}(y)$. Here
$\gamma$ is the Euler constant and $\mathrm{Ei}(x)$ the
exponential integral, so that $F(0)=0$ and $F^{\prime}(0)=-1$.
Substituting the FP value $T_c$ in
Eqs.~(\ref{rel0}) and (\ref{rel1}) we find that they are
consistent with scaling laws (\ref{scale1}) and (\ref{scale2}).
The corresponding non-universal amplitudes to one loop order  are given by
$A_R= K_d^{-1}(-iT_c)(1-\gamma T_c/2)(2\lambda\mu^2)^{-T_c/2}$,
$A_C= i\lambda A_R $, and the universal (apart from the normalization)
scaling functions to the same order read
\begin{eqnarray}
 f_R(u,v)&=& e^{-u}, \\
 f_C(u,v)&=& (v-1)\{ e^{-u}-e^{-u(v+1)/(v-1)} \notag \\
 &\times & [1-\varepsilon F(2u/(v-1))/2]  \}/u.
\end{eqnarray}
Let us introduce the FDR for a particular mode $q$ as \cite{calabrese-02}
 $X_q^{-1}=\partial_sC_q(t,s)/i\lambda R_q(t,s)$.
Using Eqs.~(\ref{rel0}) and (\ref{rel1}) we obtain to one loop order that
$X_q^{-1}= f_X (2\lambda q^2 s)$ with
\begin{eqnarray}
f_X(u)&=& 1+e^{-u} +\varepsilon [(n-3)(1-e^{-u})/(n-2)u
\notag \\
&-& e^{-u}(F(u)-2F'(u))]/2.
\end{eqnarray}
According to arguments of Ref.~\cite{calabrese-02} the
local FDR $X^{\infty}$ is identical to the global FDR
$X_{q=0}^{\infty}=\lim_{t,s \to \infty}X_{q=0}(t,s)$, which is given by
\begin{equation}
X_{q=0}^{\infty}=\frac12(1+\frac{1}4 \frac{n-1}{n-2}\varepsilon)
 +O(\varepsilon^2). \label{X}
\end{equation}
Note that $X^{\infty}_{q=0}>\frac12$ while for
most systems, except for some urn models \cite{godreche-01}, the aging
from a disordered state is characterized by the FDR $ X^{\infty}\in[0,\frac12]$.
Thus the system aging from an ordered state is more
close to equilibrium ($X=1$) than when it relaxes from a disordered state.
We expect that this applies not only to the system under consideration but also to
all systems with pure relaxation dynamics including the Ising model.

We are now in a position to discuss the relaxation of the magnetization.
The asymptotic long-time behavior of the magnetization
can be deduced  directly from Eq.~(\ref{scaling}) for $G_{10}^{00}$ and
reads $M(t)\propto t^{-\beta/\nu z}$.
To describe the behavior of the magnetization during the macroscopically early initial
stage of relaxation we derive the equation of motion for the magnetization.
The latter can be written in the form of a Ward identity
$\int_0^{\infty}ds\ \Gamma_{q=0}(t,s)M(s)=0$, which follows from
the invariance of the generating functional  under simultaneous rotation of fields
$\mathbf{s}$ and $\hat{\mathbf{s}}$ \cite{zinn-justin,bausch-80}.
Here $\Gamma_{q}(t,s)$  is a vertex function which can be expressed
as a sum of one-particle irreducible diagrams with amputated external
lines $\bm{\pi}$ and $\hat{\bm{\pi}}$.
To one-loop order we have $\Gamma_{q}(t,s)=iT^{-1}\delta(t-s)\mathcal{D}_s(q)$
with the renormalized operator
$\mathcal{D}_t=\hat{Z}Z_T^{-1}\mathring{\mathcal{D}}_t$ given for $q=0$ by
\begin{equation}
\mathcal{D}_t(0)=\partial_t+\frac{T}2\left[\gamma+\ln 2\lambda\mu^2 t
 \right]\partial_t +\frac{n-1}{4}\frac{T}{t}+O(T^2). \label{D}
\end{equation}
The equation of motion can be written  as
$ \mathcal{D}_t(0) M(t)=0 $, $t>0$ and $T=T_c$.
Solving this equation to first order in $\varepsilon$ we have to
neglect the second time derivative in Eq.~(\ref{D}) since it carries
a factor $T$ and hence in the equation of motion only gives rise to a correction
of order  $\varepsilon^2$. Having done that, we get  $M(t)=\bar{M}t^{-\varepsilon(n-1)/4(n-2)}$,
where the non-universal constant $\bar{M}$ is determined by motion of the system
during the microscopically early initial stage of relaxation.
Indeed, the considered field theory describes the  behavior of a
ferromagnet only on scales larger than the lattice constant
and on times larger than the corresponding  microscopic time  $\tau_{\text{mic}}$.
Therefore, the derived power-law behavior, which is usually called the non-linear critical
relaxation, does exist only for $\tau_{\text{mic}}<t<\tau_{\text{mac}}$.
Here $\tau_{\text{mac}}$ is the macroscopic time
at which there is a crossover to linear relaxation $M(t)\propto \exp(-t/\tau_{\text{mac}})$.
It is determined either by size of the
system $\tau_{\text{mac}}\propto L ^{z}$ or by temperature
$\tau_{\text{mac}}\propto|\tau|^{-z\nu}$.

Let us show how  this picture can be generalized to higher orders.
The vertex function is related to the corresponding
Green function by the Dyson equation \cite{zinn-justin}
$\int_0^{\infty}dt'\Gamma_{q}(t,t')R_q(t',s)(s)=\delta(t-s)$,
which is written in the time domain and can be easily check to one loop by means of
Eqs.~(\ref{rel0}) and (\ref{D}). Using Eq.~(\ref{scale1}) we
obtain
\begin{eqnarray}
&&\!\!\!\!\!\! \Gamma_{q=0}(t,s)=A_R^{-1} t^{(z-2+\eta)/z} \{[\delta(t-s)
\notag \\
&&-\left.\left. \left({t}/{s}\right)^{\bar{\theta}}
f'\left({t}/{s}\right)/s\right]\left(\partial_s-{\bar{\theta}}/{s}\right)+
\left({t}/{s}\right)^{\bar{\theta}} \psi\left({t}/{s}\right)/{s^2}\right\}, \notag
\end{eqnarray}
where $f(v)\equiv f_R(0,v)$ and  $\psi$ is the solution of Voltera equation
$\int_1^v du [f'(u)f'(v/u)v/u-f(v/u)\psi(u)]=0$ such that $\psi(\infty)=0$.
This implies that the solution of the equation of motion can be written
as $M(t)=\bar{M}t^{\bar{\theta}}f_M(t)$ with
$f_M(t)$ satisfying the condition $f_M(\infty)=1$. Taking into account
the known asymptotic behavior for $t\to \infty $ we conclude that
the identity $\bar{\theta}=-\beta/\nu z$ holds to all orders.
To find the exact form of $f_M$ we have to solve the corresponding
integral equation. Recently, the local scale invariance (LSI) was used to
predict the exact scaling
form of the response function for a system evolving from a completely disordered
state \cite{henkel-01}. We expect that the transverse fluctuations
of the considered model share similar scaling properties.
In analogy with Ref.~\cite{henkel-01} we may expect that $f(v)=1$
and consequently $\psi(v)=0$ and $f_M(t)=1$ to all orders.
However, the analysis of two loop graphs suggests that
in accordance with Ref.~\cite{pleimling-05}  there should be
small corrections to LSI: $f(v)=1+O(\varepsilon^2)$.
They come from the non-local in time contributions to the
vertex function $\Gamma_q(t,s)$
and give rise to very small corrections in the magnetization
$f_M(t)=1+O(\varepsilon^4)$ that are likely difficult
to observe in simulations.

We now discuss the relation between transverse and longitudinal
fluctuations. Let us define the longitudinal response function as
$\mathcal{R}_q(t,s)=\int d^dx\ e^{iqx} \left< \sigma(x,t)\hat{\sigma}(0,s)\right>_{\text{c}}$
and analogously  the longitudinal correlation function $\mathcal{C}_q(t,s)$.
Expressing $\sigma$ and $\hat{\sigma}$ in terms of $\bm{\pi}$ and
$\hat{\bm{\pi}}$  we obtain their low temperature expansions.
To lowest order the renormalized longitudinal response function is given by
\begin{eqnarray}
&&\!\!\!\!\!\!\!\!\!\!\!\!\!\!\!\! \mathcal{R}_{q}(t,s)  = \frac{n-1}{2 K_d} T  \left[ \ln \frac{t}{t-s}
  +   F \left({{\lambda}q^2(t-s)^2}/{2t}\right)  \right. \notag \\
&&\ \ \  - \left. F\left({\lambda}q^2(t-s)/2\right)  \right]R_q^{0}(t,s)
 + O(T^2\varepsilon, T^3). \label{R2}
\end{eqnarray}
The corresponding expression for the longitudinal correlator is too
cumbersome so that for the sake of conciseness we write down
only its time derivative at $q=0$
\begin{equation}
\partial_s \mathcal{C}_{q=0}(t,s) = \frac{n-1}{2 K_d} T^2 \lambda \ln \frac{t+s}{t-s}
+ O(T^2\varepsilon, T^3), \label{C2} \\
\end{equation}
which is needed to compute the longitudinal FDR
$\mathcal{X}_q=i\lambda \mathcal{R}_q(t,s)/\partial_s\mathcal{C}_q(t,s)$.
For the latter we obtain $\mathcal{X}_{q=0}^{\infty}=\frac12+O(\varepsilon)$.
The computation to order $\varepsilon$ requires  two-loop calculations and
is left for future investigations. However, we expect that the
scaling ansatzs (\ref{scale1})  and (\ref{scale2}) are valid also for the
longitudinal functions.

After we submitted our preprint to arxiv.org there appeared Refs.~\cite{annibale-05}
and \cite{garriga-05}  which also study the critical aging from a magnetized state
but for infinite range models such as the spherical model, long range ferromagnetic model
and Ising model in the limit of large dimension. These works confirm that the critical
aging from an ordered state is characterized by $X^{\infty}>1/2$. The $d-2$ expansions of
the response and correlation functions obtained in Ref.~\cite{annibale-05} for the spherical model
are in agreement (up to prefactors) with our results for the longitudinal functions
(\ref{R2}) and (\ref{C2}). The difference between the spherical limit ($n\to \infty$) of
the transverse FDR (\ref{X}) and the FDR of the spherical model \cite{annibale-05} suggests
that $\mathcal{X}^{\infty}\neq {X}^{\infty}$ for finite $n$ and thus one cannot
introduce the observable-independent effective temperature $T_{\mathrm{eff}}=T/{X}^{\infty}$
\cite{cugliandolo-97}.
Rewriting the response function of the spherical  model in the form (\ref{scale1})
we obtain $f_{R}^{\mathrm{sp}}(0,v)=1-(1-1/v)^{(d-2)/2}$.  Thus in contrast to the transverse
response of the $O(n)$ system with the scaling function $f_R(0,v)$  being finite in the limit
$v\to \infty$, the scaling function of the spherical model decays as
$f_{R}^{\mathrm{sp}}(0,v)\propto v^{-1}$. Supposing the same asymptotics for the longitudinal
function $f_{\mathcal{R}}(0,v)$ of the $O(n)$ system we arrive at
$\mathcal{R}_{q=0}(t,s)\propto s^{-(z-2+\eta)/z}(t/s)^{\bar{\theta}-1}$ for $t\gg s$.
This, of course, has to be checked independently.

In summary, we have studied the nonequilibrium critical behavior of
a ferromagnetic systems which evolves from a completely
ordered state. We have shown that the universal long-time scaling behavior of
magnetization emerges in the macroscopically early initial stage of relaxation
and that can be considered as a consequence of the LSI.
The two-times functions  exhibit aging behavior
which in contrast to the aging from a disordered state is characterized by
the FDR larger than $1/2$. Numerical simulations of  Ref.~\cite{zheng-96} show that
the generalized scaling behavior exists in the short-time
regime of the critical relaxation even for an initial state
with arbitrary magnetization. However, instead of the
critical exponent $\theta$  there exists a whole universal
characteristic function which up to now can be estimated only numerically.
We expect that our computation being extended to finite $h_0$
can describe this scaling behavior and provide a way to calculate
the corresponding characteristic function.
We hope that the considered model can also be useful to study
the influence of Goldstone modes on phase ordering kinetics at $T<T_c$.

We would  like to thank G.~Schehr, A.~Gambassi, M.~Henkel, B.~Zheng,
K.J.~Wiese for useful discussions and especially S.~Stepanow and M.~Pleimling
for a critical reading of the manuscript. The support from the DFG
 (SFB 418) is gratefully acknowledged.

\end{document}